\def\G{{\bf G}}

\def\be{\begin{equation} }
\def\beaa{\begin{eqnarray*} }
\def\eeaa{\end{eqnarray*} }
\def\bea{\begin{eqnarray} }
\def\eea{\end{eqnarray} }
\def\ee{\end{equation} }
\def\eq{\ =\ }

\def\etal{{\it et.al.\ }\/}

 \documentclass[final,1p,times]{elsarticle}
 \usepackage{graphics}
 \usepackage{graphicx}
 \usepackage{epsfig}
 \usepackage{amssymb}

 \biboptions{comma,square,sort&compress}
\journal{Physica E}
\begin{document}

\begin{frontmatter}

\title{\bf A real-space study of random extended defects in solids : application to disordered Stone-Wales defects in graphene.}

\author[label1]{Suman Chowdhury}
\author[label2]{Santu Baidya}
\author[label2]{Dhani Nafday}
\author[label4]{Soumyajyoti Haldar}
\author[label3]{Mukul Kabir}
\author[label4]{Biplab Sanyal}
\author[label2]{Tanusri Saha-Dasgupta}
\author[label1]{Debnarayan Jana}
\author[label2]{Abhijit Mookerjee}

\address[label1]{Department of Physics,  University of Calcutta,
 92 Acharya Prafulla Chandra Road, Kolkata 700009,India}
\address[label2]{Department of Condensed Matter and Materials Science,  S. N. Bose National Centre for Basic Sciences,Block JD, Sector III, Salt Lake, Kolkata 700098, India}
\address[label3]{Department of Physics, Indian Institute of Science Education and Research, Pune Sai Trinity Building,
Pashan, Pune 411021, India}
\address[label4]{Department of Physics and Astronomy, University of Uppsala, Uppsala, Sweden}

\begin{abstract}
We propose here a first-principles, parameter free, real space method for the study
of  disordered extended defects in solids. We shall  
illustrate the power of the technique with an 
application to graphene sheets
 with randomly placed Stone-Wales defects and shall examine the signature of such random 
defects on the density of states as a function of their concentration.
The technique  is general enough to be applied to a whole class of systems with lattice translational symmetry
 broken not only locally but by extended defects and defect clusters.
 The real space approach will allow us to distinguish signatures of specific defects
 and defect clusters.
\end{abstract}

\end{frontmatter}


\section{Introduction}
The effect of random defects on the properties of solids is an important 
area in the study of materials.  
 Defects are ubiquitous in  solids \cite{w15}. Either formed naturally during
 their preparation or artificially engineered, they may profoundly affect their physical properties. Chemical reactions, phase transitions or plastic deformations during the formative stage  may leave their imprints as defects. These defects may be local, like vacancies \cite{w17,w16},
 substitutional impurities \cite{w18} or  adatoms \cite{w19}.
Defects can also be extended, like dislocations, stacking faults, twins and grain boundaries. 
Extended defects have been visualized by STM images \cite{guo}.  It is important therefore to set up
a first-principles, essentially parameter free, theoretical methodology for the signature of random defects, 
and to distinguish, in particular,  the role of disorder. Local random defects has been thoroughly studied using sophisticated mean-field approaches like the itinerant coherent potential approximation (ICPA) \cite{icpa}, the travelling cluster approximation (TCA)\cite{tca}, both derived from the parent formalism : 
 the augmented space formalism (ASF) \cite{asf2,asr2}. However, most of the work on extended defects have involved super-cell methods. These works look at essentially periodic arrays of defects with disorder
stretching only over the finite super-cell \cite{ahuja,popov,waghmare}. Long ranged disorder is not accessible to these reciprocal space based methods. None of the  supercell methods can accurately capture the disorder
 induced smearing of the density of states due to  the self-energy arising out of scattering of Bloch
 waves by configuration fluctuations. This `life-time' effect is experimentally accessible through neutron
 scattering and has been found to be strongly dependent both on the energy E and the wave-vector $\vec{k}$. In this work we shall propose a fully real space technique, where the structure and Hamiltonian of the relaxed defected lattice
goes as an input into a recursive algorithm which gives us the electronic structure carrying the signatures of disorder. Although the application will be to a specific problem, the method is general enough to be applied in any solid carrying random extended defects. 

Electronic structure of low-dimensional solids  with extended defects have been addressed earlier \cite{ahuja, popov, waghmare}. Let us quote 
Shirodkar and Waghmare \cite{waghmare} :
\begin{quote}
``Our work, along with other work \cite{ahuja,popov} has involved a {\sl periodic array} of SW defects,
whereas SW defects are {\sl randomly distributed} in a real sample."
\end{quote}
  The authors go on to remark that although this disorder may not affect the vibrational spectrum significantly, 
the electronic structure obtained from a periodic array of SW defects needs to be interpreted with care. Our real 
space recursion approach \cite{heine} makes no appeal to lattice translational symmetry and is thus ideal for studying 
topological  disorder and its effect on  electronic structure. The  illustration of this point is the justification of our 
proposed methodology and the main focus of this communication.

\section{The real space recursive algorithm}

Early in the seventies Heine and co-workers introduced \cite{md27,heine} the recursion method based on a fully real space
technique to deal with lattices without any translation symmetry. The essential inputs are the geometry of
the lattice and a tight-binding, sparse Hamiltonian.  

\begin{figure}[b]
\centering
{\framebox{\includegraphics[width=3.cm,height=3.cm]{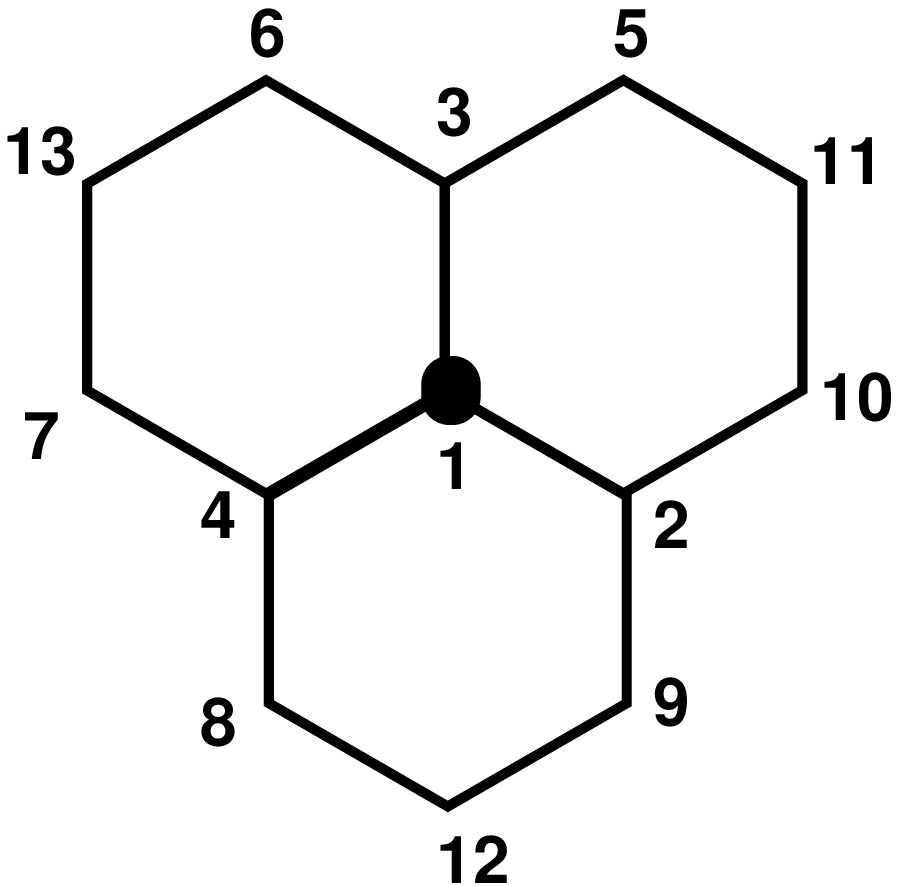}}}\hskip 1cm
{\framebox{\includegraphics[width=3.cm,height=3.cm]{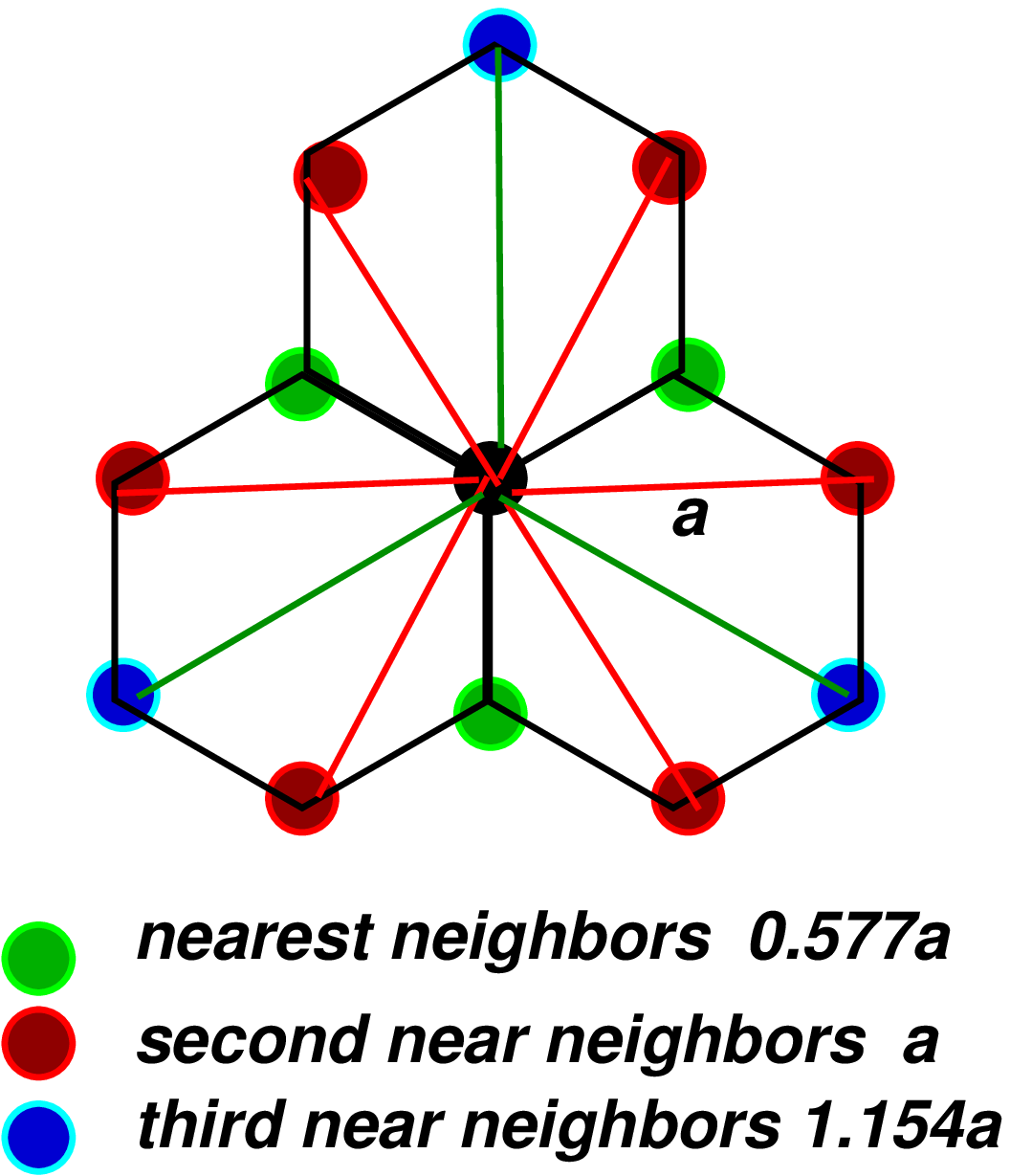}}}\hskip 1cm
{\framebox{\includegraphics[width=3cm,height=3.2cm]{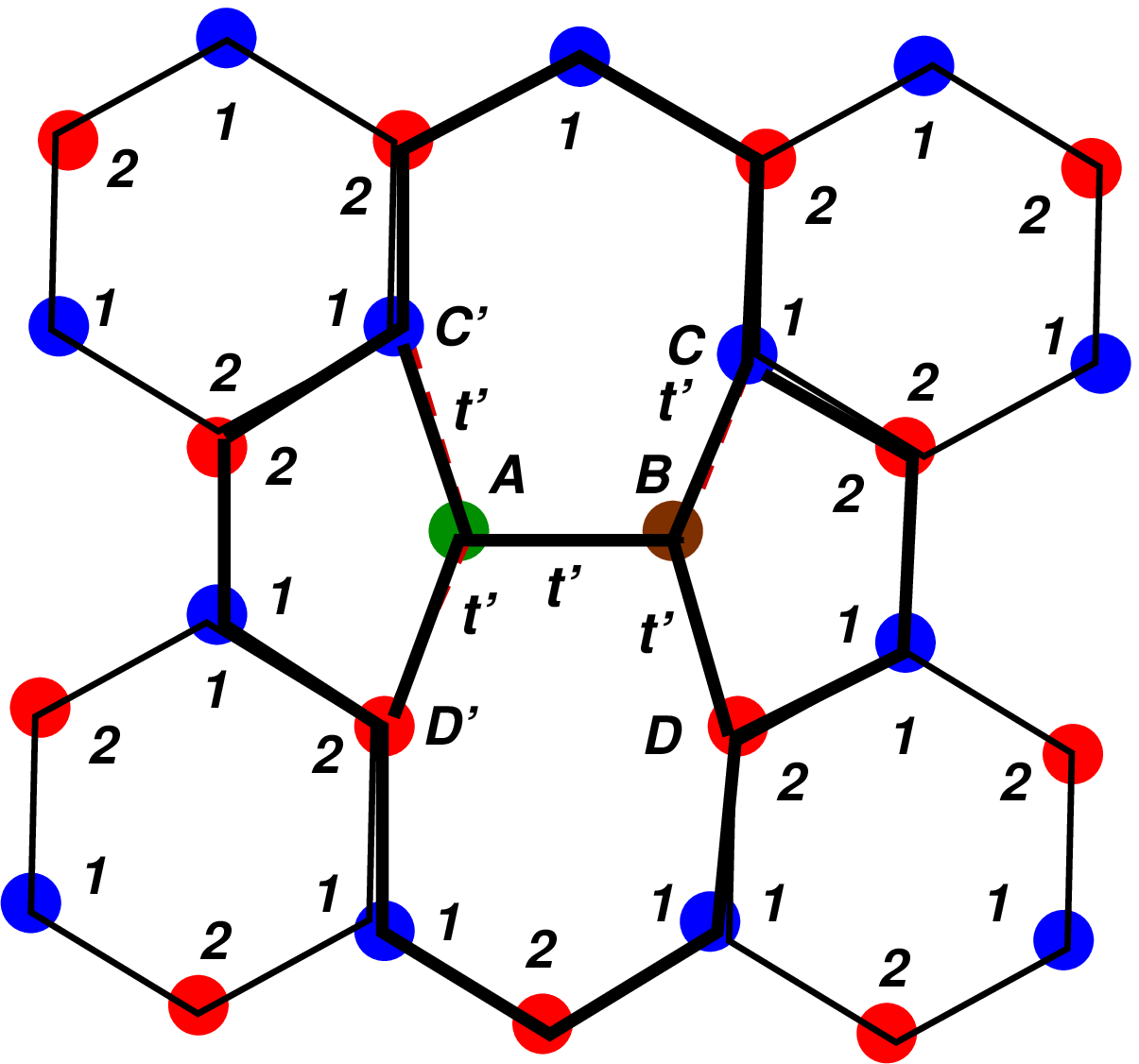}}}
\caption{(Colour online) (left panel) 
The  mapping of lattice sites on a honeycomb lattice onto a denumerable set of positive integers.
(center) The nearest, next-nearest and third nearest neighbour vectors on a 
honeycomb  lattice.
(right panel) Lattice distortions and change in Hamiltonian elements around a structural defect. \label{fig5}}
\end{figure}

As an example, and in view of our later applications, let us look at a honeycomb lattice and map its vertices
onto the set of positive integers. The mapping is not unique, but necessarily one-to-one (Fig.\ref{fig5}). The geometry of
the lattice is uniquely described by a connectivity matrix $C(n,m)$ which gives us the $m$-th neighbour of
the vertex labelled as $n$. The central panel of Fig.\ref{fig5} shows a model in which up to third
neighbour overlaps are taken into account. The connectivity matrix is :

\be C(n,m) =  \left(\begin{array}{c|ccc|cccccc|ccc}  1 & 2 & 3 & 4 &5 &6 &7 &8 & 9 & 10& 11&12&13\\ 2 & 9 & 10 & 1&3&4&12& 11&\ldots&\ldots &5&8&\ldots \\ 3& 6&5 &1&11&4&2 &13 &\ldots&\ldots &7&10&\ldots \\ 4&7 &8 &1  & 3& 12&2&13&\ldots&\ldots&6&9 &\ldots  \\ \multicolumn{4}{c}{\ldots\ldots\ldots}\end{array} \right)\label{map}\ee

The tight-binding Hamiltonian can be written as :

\be H = \sum_n\sum_\alpha \epsilon_\alpha P_{n\alpha} + \sum_{n\ne m}\sum_{\alpha\beta} t_{\alpha\beta}(n-m) T_{n\alpha,m\beta} \ee

Here,  ${P}$ and ${T}$ are projection and transfer operators on the tight-binding basis labelled by $n,\alpha$, where $n$
is the vertex label and $\alpha$ all other possible degrees of freedom associated with it.
Similarly, we describe a honeycomb lattice with a structural defect shown in the right most panel of Fig.\ref{fig5}. We label the distorted sites say with a label $\gamma$, then whenever $\alpha,\beta\not=\gamma$, $t_{\alpha\beta}(n-m)\eq t(n-m)$ and
otherwise $t_{\alpha\beta}(n-m) \eq t'(n-m)$.

With  this recursion method, we shall study the representations of the Green operator or the resolvent of the Hamiltonian,

\[ {\bf G}(z) = (z{\bf I}-{\bf H})^{-1}\]
where $z$ is a complex variable.  Most physically measurable quantities describing the  electronic properties 
of a solid  are related to different  matrix elements of {\bf G}(z). In particular, the  local (atom
projected) density of 
states (LDOS) $n_i(E)$ and the total densities of states (TDOS) $n(E)$ :

\begin{eqnarray}
-\frac{1}{\pi}\lim_{z\rightarrow E-\imath 0^{+}}\Im m\ G_{ii}(z)& =&  n_i(E)\nonumber\\
-\frac{1}{\pi}\lim_{z\rightarrow E-\imath 0^{+}}\frac{1}{N}\Im m\ {\rm Tr}\ \G(z)&\eq &n(E)
\end{eqnarray}

Where N is the number of atoms in the system. It would be interesting to note that although the super-cell
based methods can easily access both the band projected partial density of states (PDOS) and the total 
density of states (TDOS), it is only the real-space based methods that can give is a single atom or an
atomic cluster projected {\sl local} densities of states (LDOS). 

   In this
basis the representation of the Hamiltonian  is a matrix of infinite rank. 
The solution of the Kohn-Sham equation of an electron in this system 
 can be simplified enormously if the Hamiltonian has lattice translation symmetry. 
  In that case the Bloch theorem introduces the quantum  label $\vec{k}$ and  reduces the
 effectively infinite rank matrix representation of the Hamiltonian to a manageable finite rank equal to
 the number of bands (in this case two). 
Consequently, crystalline systems are always studied in this Bloch representation. But in disordered systems, particularly
where the disorder is a topological distortion of the lattice and these defects are distributed randomly throughout it,   
such periodicity is absent and the reciprocal space representation is no longer strictly valid.
In such situations we need to look for real-space based techniques. 

Calculation of the resolvent requires inversion of a matrix of infinite rank. 
Haydock \etal \cite{md27} proposed a technique to do so. We start from a suitable vertex  at $|1\rangle$.
It  essentially involves generation of a new basis through a three term recurrence relation :

`\begin{eqnarray}  \vert 1\} & \eq & \vert n\alpha\rangle \nonumber\\
  \vert 2\} & \eq & (H-\alpha_1)\vert 1\}  \nonumber\\
  \vert n+1\} & \eq & (H-\alpha_n)\vert n\} - \beta_{n+1}^2 \vert n-1\} 
\end{eqnarray}

\be \alpha_n \eq \frac{ \{n\vert H\vert n\}}{\{n|n\}} \quad \mbox{and}\quad \beta_n^2 \eq \frac{\{n\vert n\}}{\{n-1|n-1\}} \ee

Haydock \etal\cite{md27} showed that the matrix element of the resolvent may be written as a continued fraction.

\begin{eqnarray}
\langle n\alpha\vert G(z)\vert n\alpha\rangle & = & \{1\vert G(z)\vert 1\} 
 =  \frac{1}{\displaystyle z-\alpha_1-\frac{\displaystyle \beta_2^{2}}{\displaystyle z-\alpha_2-\frac{\displaystyle\beta_3^{2}}{\displaystyle z-\alpha_3-\frac{\displaystyle \beta_4^{2}}{\displaystyle \ddots z-\alpha_N-\beta_N^2 T(z)}}}}\nonumber\\
\end{eqnarray}

Right at the start we emphasized that we have chosen a real space algorithm over mean-field and supercell approaches, because we do not wish to introduce artificial periodicity and confine randomness over a  finite volume. But the problem with any  
 numerical calculation  is that we can  deal with only a finite number of operations.  In the recursion algorithm, we can go up to a finite number of steps leading to exactly what we wish to avoid. The analysis of the asymptotic part of the continued fraction then is of prime interest to us.    This is the ``termination" procedure in which the asymptotic behaviour is assessed from the way the coefficients $\{\alpha_n,\beta_n\}$ 
behave as $n\rightarrow\infty$. Different terminators  have been discussed in detail by Haydock and Nex \cite{ter1}, Luchini and Nex \cite{ter3}, Beer and Pettifor \cite{ter4} and in considerable depth by Viswanath and Muller \cite{ter5}. The terminator which describes the asymptotically far environment must satisfy certain basic properties : 
This termination process is a delicate mathematical approximation. We would like to maintain the Herglotz analytic properties of the resolvent after  termination. 
A function T(z) is called Herglotz if : 
\begin{enumerate}
\item[(i)] All singularities of T(z) lie on the real z-axis.
\item[(ii)] Singularities of T(z) form a bounded set.
\item[(iii)] Im\ T(z)$\geq$ 0 if Im\ z $<$ 0 ; Im\ T(z) $\leq$ 0 if Im z $>$ 0.
\item[(iv)] Re T(z) $\rightarrow$ 0 as Re z $\rightarrow \ \pm\infty$
\end{enumerate}

The next step is to analyze our resolvent to locate singularities on its compact spectrum. Majority of resolvents with bounded spectra have singularities at the band edges.
The termination of continued fractions describing spectral densities with
compact support and singularities except at the band edges have been 
described in detail in earlier works \cite{ter1,ter3,ter4,ter5}.  

The best illustration of this technique is to apply it to an interesting material where alternative methods 
always leads to problems of one sort or another. We have chosen to study graphene with random Stone-Wales defects.

\section{The relaxed graphene lattice with random Stone-Wales defects.}

\begin{figure}
\centering
\framebox{\framebox{\includegraphics[width=6cm,height=3cm]{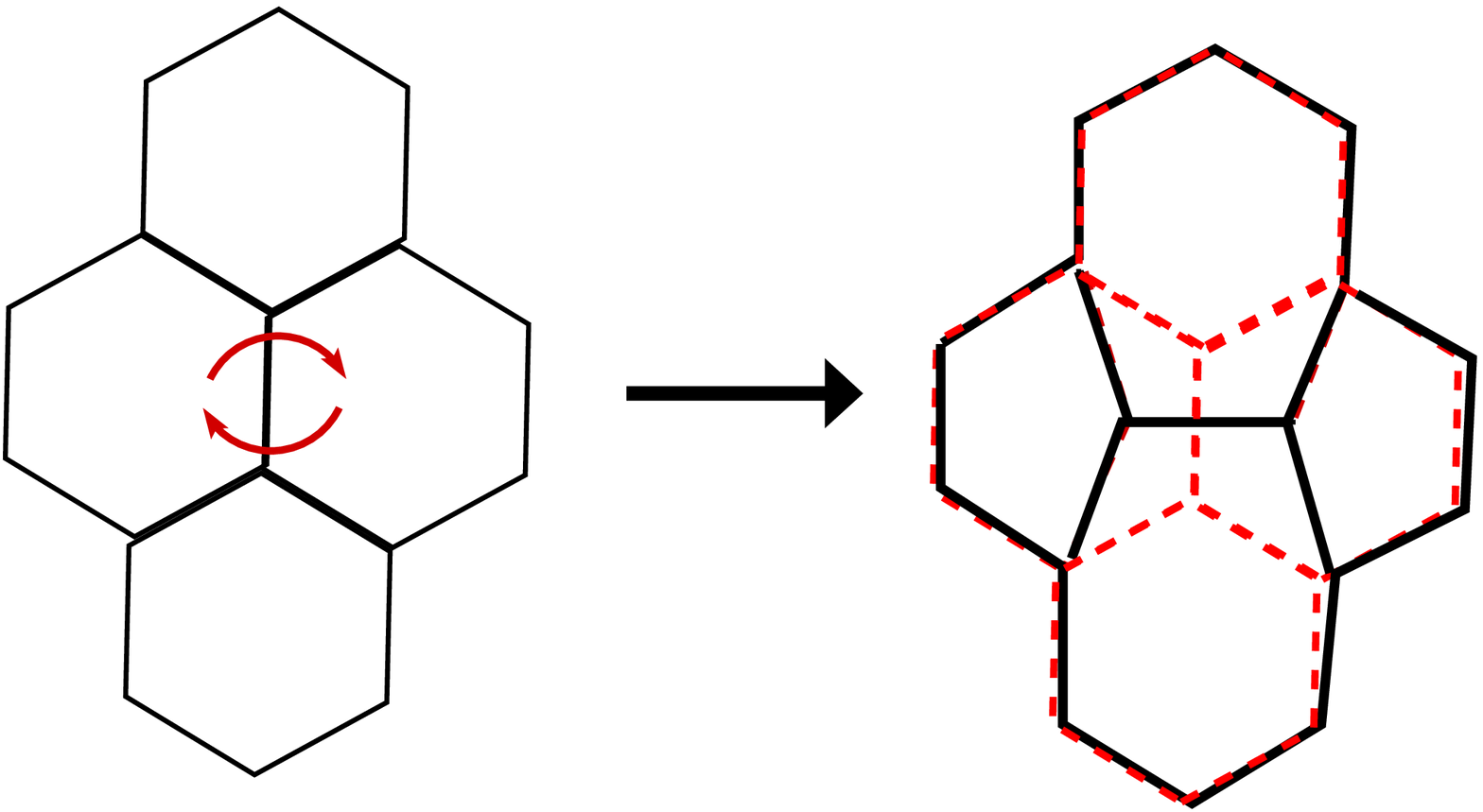}}}
\caption{(Colour online) SW defect formation due to a  $\pi/2$ rotation of a C-C bond \label{fig1}}
\end{figure}

 In this work, we shall be interested in intrinsic extended structural defects like Stone-Wales (SW) defects \cite{stonewales} in graphene.
The study of graphene, the two-dimensional allotrope of Carbon is an area of intense interest to both theoretical and experimental
 researchers for various reasons. The first is the  possible technological applications of graphene\cite{w2,md2,md1,w1,w3}. 
Second, graphene seems to defy the theorem of Mermin and Wagner \cite{mw}  and form a stable two-dimensional structure. Finally,  the 
electrons in graphene behave like massless charged particles, something not encountered in our three dimensional world \cite{fuchs}. There has even been fanciful applications of general relativistic ideas in graphene.
 The SW defects are formed by $90\,^{\circ}$ rotation of
 a C-C bond. After such a rotation, subsequent re-bonding leads to  the formation of two pentagonal and two heptagonal carbon rings.  This is illustrated in Fig.~\ref{fig1}.
 These defects are responsible not only for bringing  about fractures and embrittlement
 \cite{jie2,jie4,jie5,jie6}, but also for altering  the chemical reactivity and
 transport properties \cite{jie7,jie8,jie9} and for causing the rippling behavior of graphene sheets \cite{zol}.
 SW defects can lead to out-of-plane displacement of carbon atoms in graphene\cite{w12,jie10,jie11,w11} and induce curvature \cite{jie12}. Recently, based on these three-dimensional ripples on 
the graphene sheets with SW defects, the assumption that SW defects are perfect two dimensional  has been questioned \cite{jie15,jie8,jie13,jie7,jie16,jie17,jie14}.

\begin{figure}[b!]
\centering
{\framebox{\includegraphics[width=3.5cm,height=3.5cm]{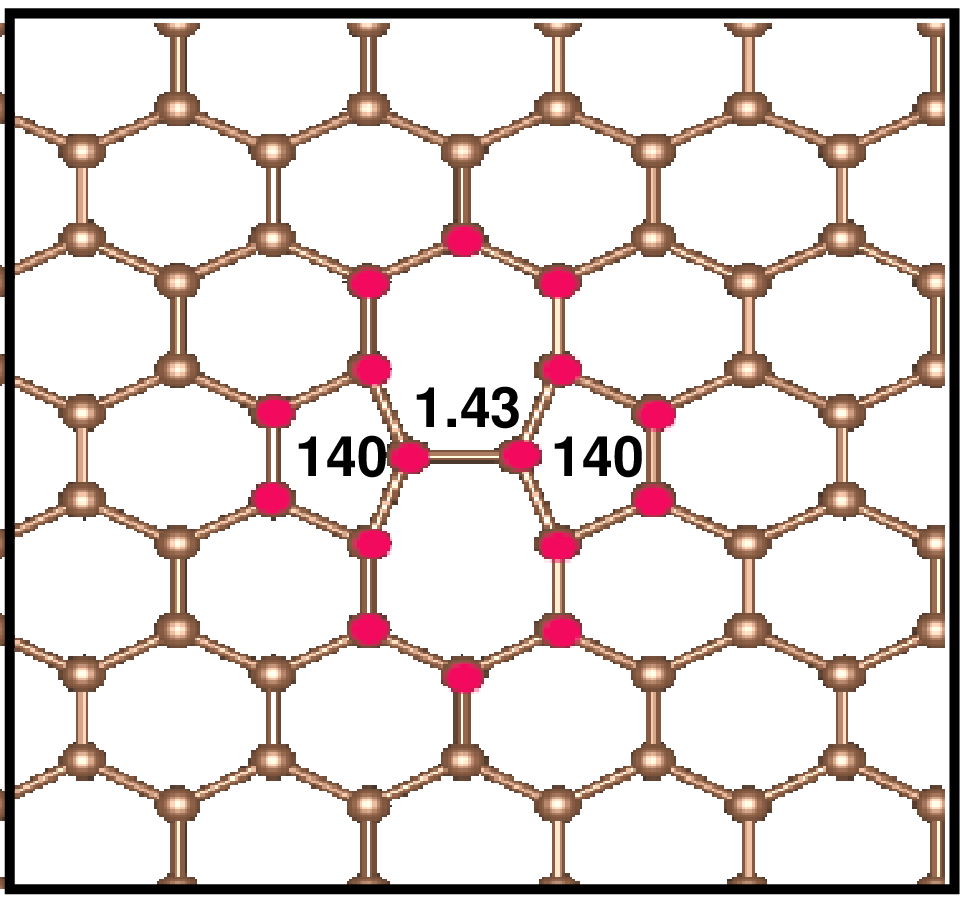}}}\hskip 2cm
{\framebox{\includegraphics[width=3.5cm,height=3.5cm]{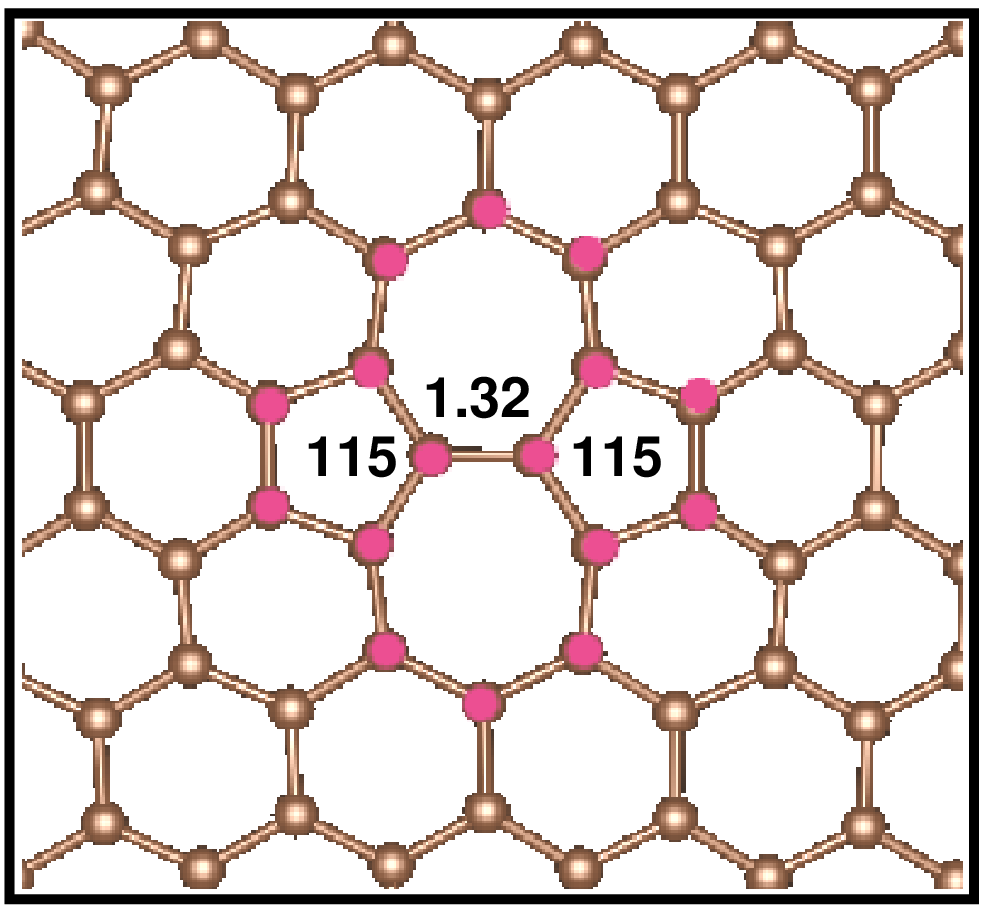}}}
\caption{(Left) A vertical SW defect in the unrelaxed lattice (Right) The same defect after planar relaxation.
\label{fig2}}
\end{figure}

For setting up the lattice model, we have started with  a flat graphene sheet cluster, containing 3200 atoms. Different concentrations of Stone-Wales defects 
were then  created 
by randomly choosing bonds on the lattice, rotating by $\pi/2$ and rebonding as shown in Fig. \ref{fig1}.
The first step of relaxations all lay on the graphene sheet and involved mainly on-plane relaxation of the C-atoms. The left panel
in Fig.\ref{fig2} 
shows a vertical SW defect in  the unrelaxed lattice while the right panel shows the effect of relaxation on the defect.

Lattice  relaxation was carried out using the 
  density functional tight binding method as implemented in a sparse-matrix based  DFTB+ code \cite{aradi}. 
In this method, a first-principles
  form of density functional  is used and hence, large systems could be handled with reasonable accuracy. 
We have used the conjugate gradient method to relax the internal coordinates with 10$^{-2}$ eV/A force convergence for
 the ionic relaxation and 10$^{-4}$ eV energy convergence for the electronic relaxation. Geometry optimizations were 
done using calculations at the $\Gamma$ point in the Brilluoin zone (BZ). The Fermi smearing method has been used with an electron temperature of 100K.  

This relaxation first causes the rotated C-C bond length to compress to 1.32A from its unrelaxed value of 1.44A. As a consequence of this shortening of bond length,the bond angle at the apex of the pentagon compresses from 140 degree to 115 degree,which is ~18.5$\%$. Because of these large changes the system experiences a compressive
 stress along the direction of unrotated C-C bonds,and tensile stress along the rotated C-C bond.The graphene sheet then reduces this in plane stress by buckling in the z-direction,  forming a non-planar, rippled structure. On structure relaxation the system energy is lowered.  
\begin{figure}[t!]
\centering
{\framebox{\includegraphics[width=5cm,height=3.cm]{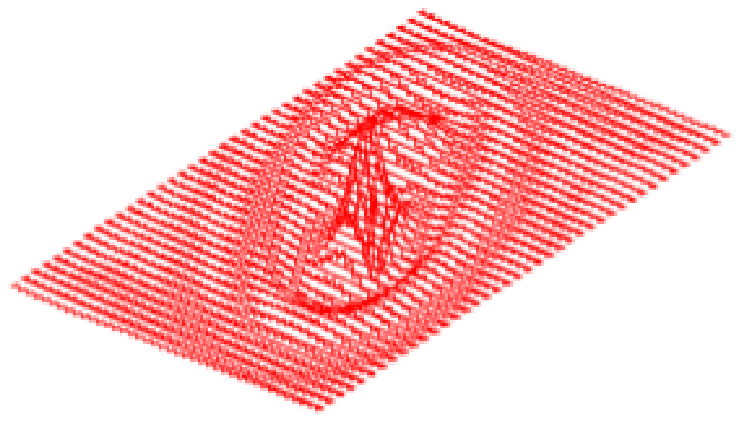}}}\hskip 1cm
{\framebox{\includegraphics[width=5cm,height=3.cm]{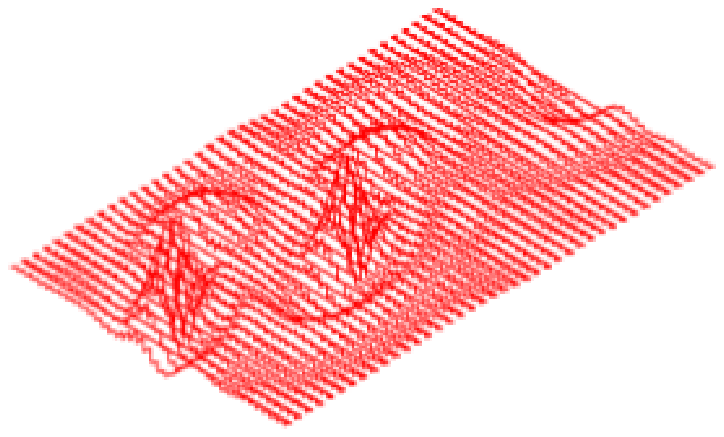}}}\\
{\framebox{\includegraphics[width=5cm,height=3.cm]{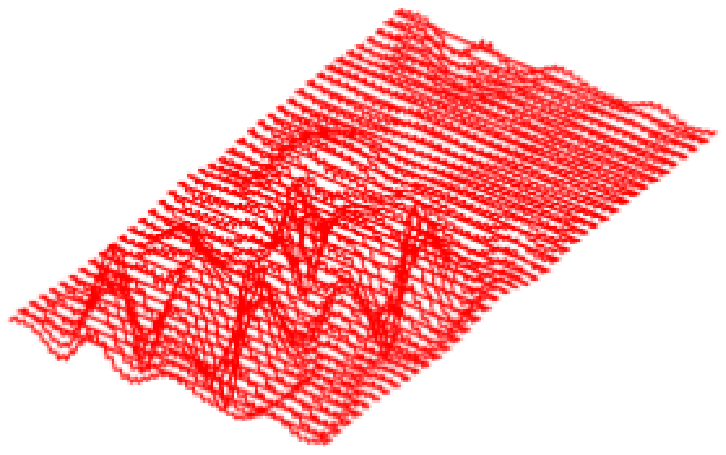}}}\hskip 1cm
{\framebox{\includegraphics[width=5cm,height=3.cm]{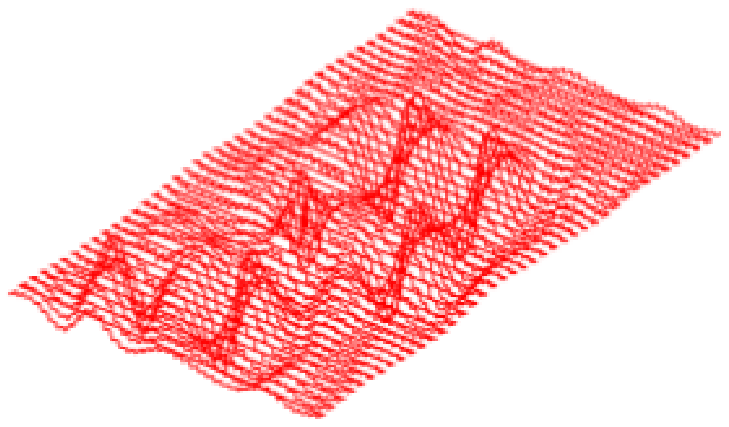}}}\\
\caption{(Top, Left) Rippling of the graphene lattice on a single SW defect. (Top, Right and Bottom) Rippling
of a graphene sheet with varying concentrations of SW-defects.}
\label{fig3}
\end{figure}

Once the in-plane relaxation was completed we went a step further and relaxed the lattice in full three dimensions, allowing
movement of atoms perpendicular to the original graphene plane. The top left panel of Fig. \ref{fig3} shows the rippling of
the graphene sheet on a single SW defect. The other three panels show rippling in a sheet with higher concentrations of SW-defect. Full lattice relaxation is thus extremely important for the structural changes on relaxation may have very important effect on the electronic structure
of the materials.  
 We shall deal with the effects of structural changes with Harrison scaling\cite{harrison1,harrison2}.

\section{First Principles derivation of the Hamiltonian.}

Once we have constructed the relaxed graphene sheets with SW defects, the next step is to obtain a Hamiltonian for the band due to the $\pi$-bonded $p_z$ electrons from first-principles, as far free from fitted parameters as possible.
The calculations  begin with a self-consistent ground state calculation for the single layer graphene using 
the tight-binding, linear muffin-tin orbitals (TBLMTO) method within  the atomic sphere approximation.
Three empty spheres were needed to achieve  space filling.
The muffin tin radii used for carbon (C) and the empty spheres were 1.56 a.u. and 2.76 a.u. respectively.
The minimal basis set for the self-consistent calculation consisted of C $s,p_x,p_y,p_z$ and empty sphere s states.
Fig.\ref{fig3} left panel shows the full TB-LMTO bands. 

\begin{figure}[b!]
\centering
{\includegraphics[width=5cm,height=4.0cm]{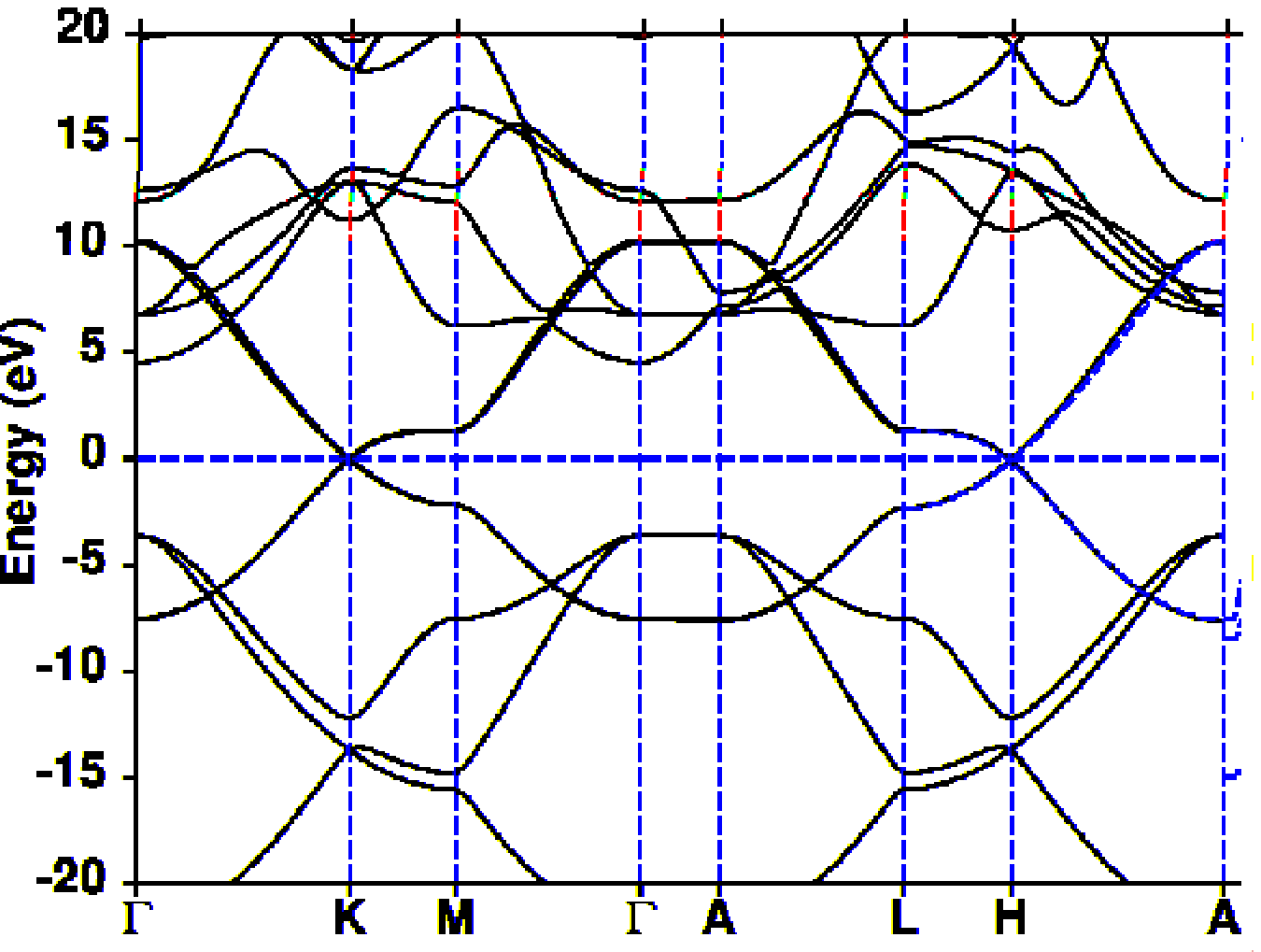}}
{\includegraphics[width=5cm,height=4.0cm]{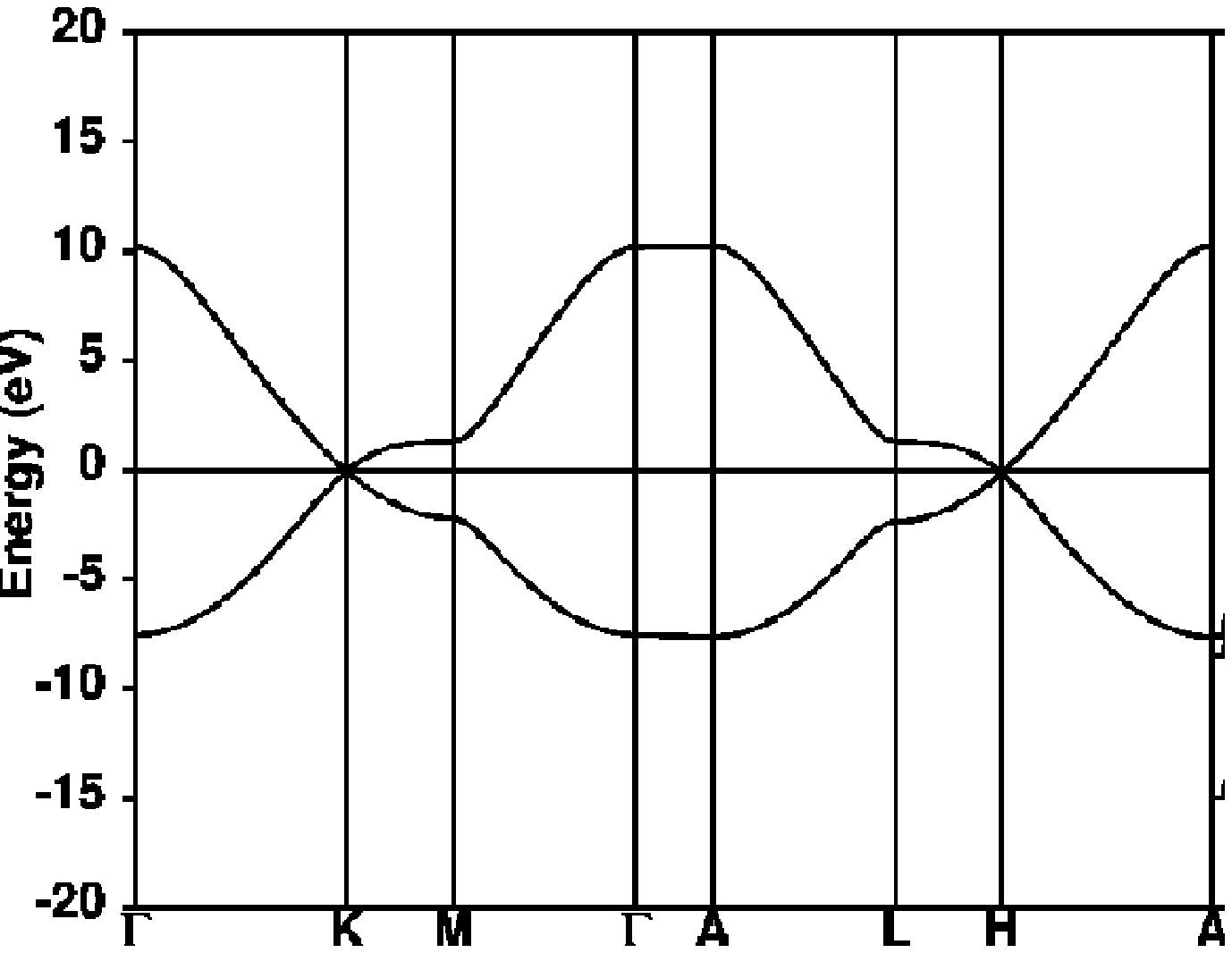}}
\caption{(Colour online) (Left panel) The TB-LMTO bands for pristine graphene (Right panel) The NMTO down-folded bands due to $\pi$ bonding between the p$_z$ states. 
\label{fig4}}
\end{figure}

The {\sl active orbitals} were recognized as the $p_z$. The remaining  degrees of freedom were integrated out \cite{asd,slv} using a  
downfolding procedure. Formally, we partition the Hilbert space ${\cal H}$ in which the Hamiltonian is defined into one which 
is spanned by the active orbitals $\cal{H}_{\rm act}$ and the rest, which we shall call the `bath' : ${\cal{H}} = {\cal {H}}_{\rm act} \otimes \cal{H}_B$ so that 
\[ H = \left( \begin{array}{c|c}
           H_{\rm act} &  H'\\ \hline
            H'     & H_B  
\end{array}\right)\]

The partition or  downfolding theorem gives the effective Hamiltonian in the subspace ${\cal {H}}_{\rm act}$ :

\[ H_{\rm eff} = H_{\rm act} + H'^\dagger\ G_B(\varepsilon_n)\ H'\]

$G_B$ is calculated by inverting $(\varepsilon_n I-H_B)$ in the bath subspace ${\cal H}_B$.
For the present study, only 
the C-p$_{z}$ orbitals were kept active. All others including C-s, p$_{x}$, p$_{y}$ and empty sphere
s-orbitals were 
downfolded. The integrated out orbitals renormalized the active C-p$_{z}$ orbital. 
The 
renormalized C-p$_{z}$ NMTOs were constructed using four energy mesh points $\varepsilon_{0}$-
$\varepsilon_{3}$ which gives 3$^{rd}$ order muffin tin orbitals. The energy mesh used for the construction of NMTOs were chosen to be in the energy window spanning the C-p$_{z}$ bands. The minimal set of C-$p_{z}$ NMTOs serve the purpose of 
effective localized C-p$_z$ Wannier functions. 

\begin{figure}[h!]
\centering
{\includegraphics[width=6cm,height=4.0cm]{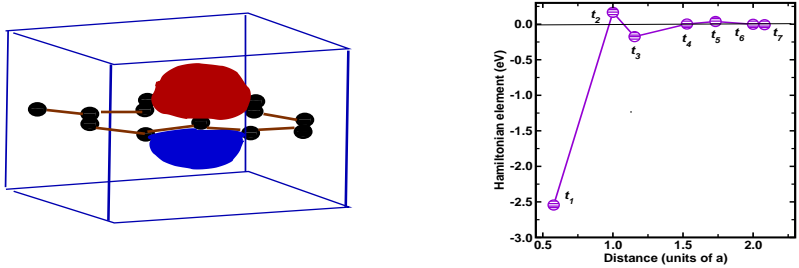}}\hskip 1cm
{\includegraphics[width=4cm,height=3.5cm]{Fig5d.eps}}
\caption{(Colour online) (Left) $p_z$ orbitals perpendicular to the graphene plane. (Right) 
Hamiltonian parameters \label{pz}}
\end{figure}

The effective C-p$_z$ -  C-p$_z$  hopping interaction:  $t(R, R^{\prime})$,  connecting C-p$_{z}$ NMTO-s 
$\chi(R)$ centered at the site ${R}$  to the neighbouring C-p$_{z}$ NMTO $\chi(R^\prime)$ centered at the $R^{\prime}$,
were obtained by Fourier transformation of the low 
energy C-p$_z$ Hamiltonian H$_{p_z}(\vec{k})$ \cite{dress,slv}. These  are shown in Fig.\ref{pz} and Table 1.
\begin{table}
\centering
\begin{tabular}{|c|cccc|}\hline
(eV) & $\varepsilon$ & t$_1$ & t$_2$ & t$_3$  \\ \hline 
& -0.291   & -2.544  & 1.668   & -1.586\\
\hline
\end{tabular}
\caption{Tight-binding parameters generated by NMTO.\label{tab1}}
\end{table}

The downfolded bands are shown on the top right panel of Fig.\ref{fig4}. The corresponding $t(R)$ is shown in the bottom  
right panel of Fig.\ref{fig4} and Table \ref{tab1}. Please note that the overlap
is reasonably long ranged. Thus some of the earlier calculations with a simple nearest neighbour overlap (Reich \etal\cite{reich}) 
 may not show the accurate picture. Longer ranged overlap models have also been attempted. However, all of them are fitting procedures to the single C-p$_z$ bands. Our procedure of getting the bands from first-principles DFT calculations and then systematically down-folding away the non-active bands is both physically more appealing and mathematically more rigorous.

\section{The Terminator}

Having obtained the one band Hamiltonian (arising from the $\pi$ bonded $p_z$ orbitals), we carry out the
recursion algorithm up to a finite number of steps.
As discussed earlier, from the asymptotic behaviour of the recursion coefficients we estimate the terminator.
We analyze our resolvent to locate singularities on its compact spectrum. Majority of resolvents with bounded spectra have singularities at the band edges.
The termination of continued fractions describing spectral densities with
compact support and singularities on it have been 
described in detail in earlier works \cite{ter1,ter3,ter4,ter5}.  
For graphene, we expect  the $\pi$ bands to have a spectral density which has an additional singularity at the Dirac
point. Terminators appropriate to such problems have been discussed by Magnus \cite{mag} and Viswanath and M\"uller \cite{ter5}. They propose a terminator of the form :
\be
T(z) = \frac{2\pi (E_m)^{(\alpha+2\beta+1)/2)}}{B\left(\frac{\alpha+1}{2},1+\beta\right)}\ 
 \vert z-E_0\vert^\alpha \left\{(z-E_1)(E_2-z)\right\}^\beta 
\ee

\begin{figure}[h]
\centering
\includegraphics[width=6cm,height=4cm]{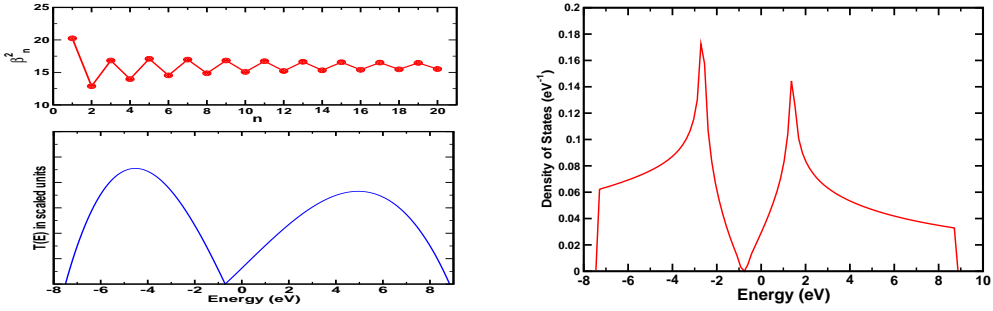}\hskip 1cm
\includegraphics[width=6cm,height=4cm]{Fig6b.eps}\\
\caption{(Left,Top) The terminator with an internal singularity appropriate for graphene.
(Left,bottom) Asymptotic parts of the calculated pristine graphene Green function continued fraction coefficients obtained by recursion. (Right) The TDOS of pristine graphene taking up to third nearest neighbour overlaps.\label{term}}
\end{figure}

In our problem, $E_0$ is the Dirac point and $E_1, E_2$ are the spectral bounds, $E_m^2=E_1E_2$, $\alpha\ =\ \beta\ =\ 1$.

Magnus \cite{mag} has cited a closed form for the continued fraction coefficients of the terminator~:

\be
\beta^2_{2n}  =  \frac{4E_m^2 n(n+1)}{(4n+2)(4n+4)} \quad 
\beta^2_{2n+1}  =  \frac{E_m^2(2n+2)(2n+4)}{(4n+4)(4n+6)}
\ee

The parameters of the terminator are estimated from the asymptotic part of
the continued fraction coefficients calculated for our problem. The top 
left panel in Fig.\ref{term} shows the continued fraction coefficients for
the Viswanath-Muller terminator shown in the bottom left panel of this figure. The calculated continued fraction coefficients from recursion are shown in the
right panel. The parameters of the terminator coefficients are fitted from these results.
The resultant density of states is shown in the right panel of Fig.\ref{term}. Unlike the usual nearest neighbour models the density is not symmetric round the Dirac point. If we
write the Green function as contributions from non-intersecting paths (based on Feenberg perturbation \cite{ter3}) then in the nearest neighbour model all non-intersecting paths are of even length and all $\{\alpha_n\}$ is a constant, leading to a symmetric density. As soon as we introduce longer ranged $t$, odd length non-intersecting paths appear. Then $\alpha_n$ vary with $n$ and the density becomes non-symmetric.
  
\begin{figure}[b!]
\centering
{\framebox{\includegraphics[width=3cm,height=3cm]{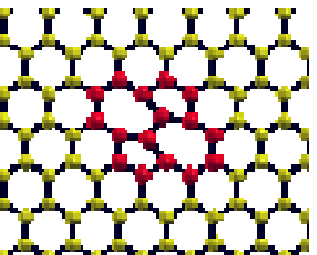}}}\hskip 0.5cm
{\framebox{\includegraphics[width=3cm,height=3cm]{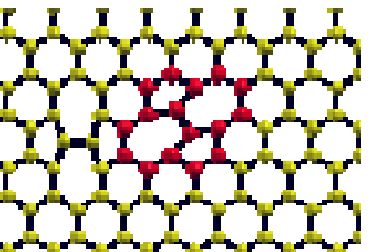}}}\hskip 0.5cm
{\framebox{\includegraphics[width=3cm,height=3cm]{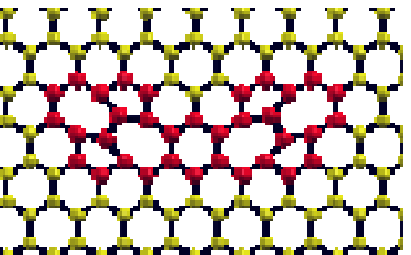}}}\\
{\framebox{\includegraphics[width=3cm,height=3cm]{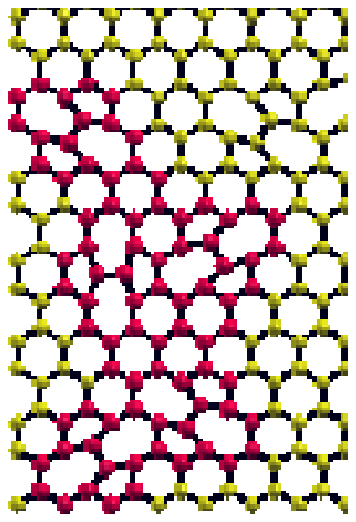}}}\hskip 0.5cm
{\framebox{\includegraphics[width=3cm,height=3cm]{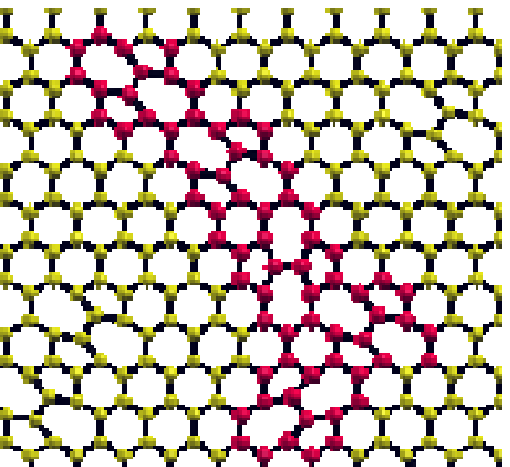}}}\hskip 0.5cm
{\framebox{\includegraphics[width=3cm,height=3cm]{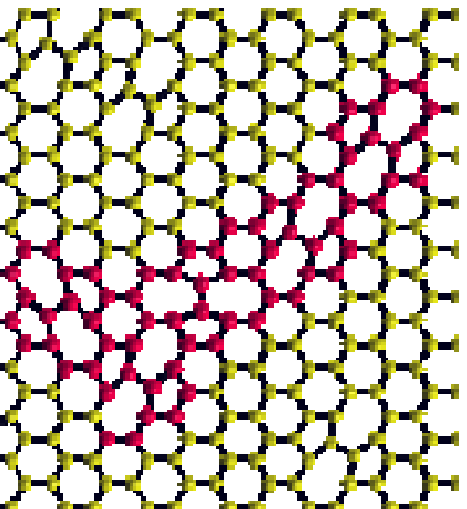}}}
\caption{(Top panel) Single and connected double defects sharing a hexagon. (Bottom) Connected
clusters of defects in systems with higher defect concentrations.\label{dos}}
\end{figure}

\section{Results and Discussion.}

\begin{figure}
\centering
{\includegraphics[width=8.cm,height=7.cm]{Fig8a.eps}}
{\includegraphics[width=5.5cm,height=6cm]{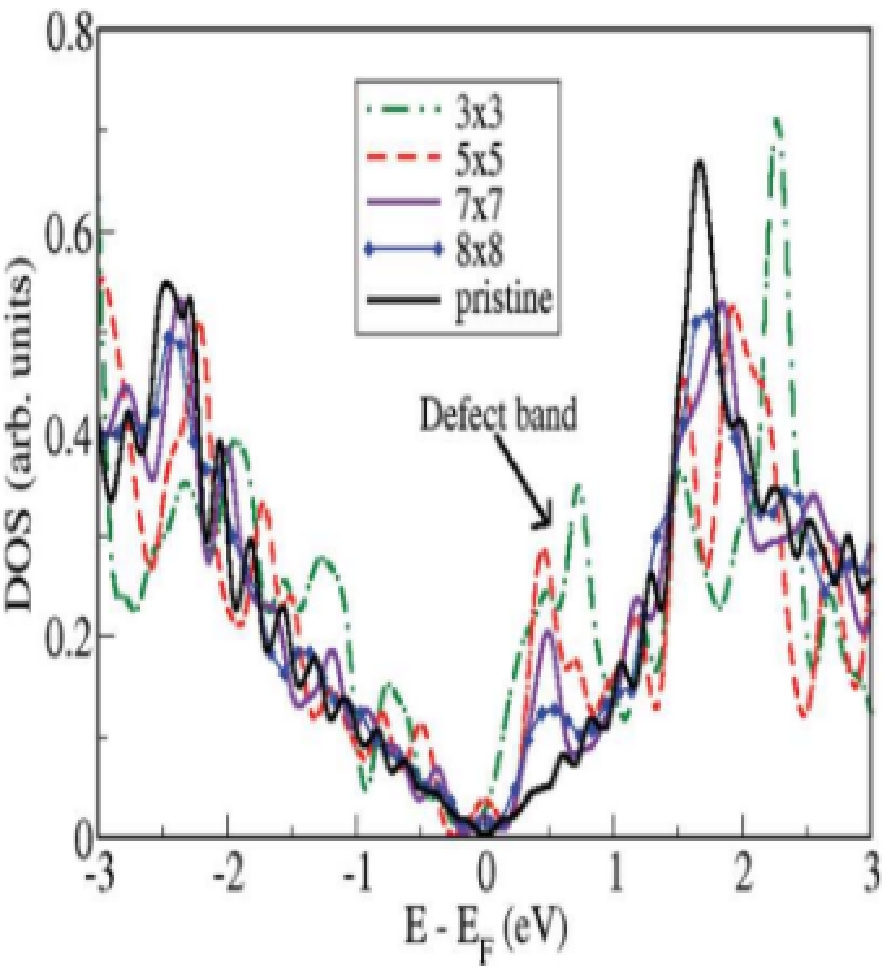}}
\caption{(Colour online) (Top) The total density of states near E$_F$ for four different SW defect concentrations. (Bottom,left) The supercell results of 
Shirodkar and Waghmare.\label{fig10}}
\end{figure}

Since the axis of the SW defects are along the bond which is rotated in its formation, we note that the defects are either vertical or at angles 60$^o$ tilted to the left and right of the vertical. Closeups of two such isolated tilted defects are shown on the top row of Fig.\ref{dos}. We also show a double defect one tilted to the left and another to the right sharing a common hexagon. In the samples with higher concentrations of defects we also have clusters of connected defects. Two such examples of defect clusters are shown in the bottom row of Fig.\ref{dos}.

In order to analyze the signature of individual defects we first show the total density of states  for several defect concentrations 
This is shown in the top panels of Fig.\ref{fig10}. We note that
 the occupied part of the spectrum below the Dirac point show very little change.  The isolated defects
 show a defect state around 1 eV above the Dirac point.  For the double defect this structure widens.
 There is also a defect induced signature around 3-4 eV above the Dirac point. These signatures were also
 observed by Shirodkar and Waghmare\cite{waghmare} (compare with their Fig. 5).  Of course, in our real space
 picture there is no concept of defect `bands', but the spectral signatures are very similar. 

 The supercell approaches  indicate the local defect
levels are very similar to our real space approach. However it has been known that the effect of
the far disordered environment leads to a 'self-energy' \cite{w15}, the
real part of which shifts the defect levels and the imaginary part causes them to widen. In our continued fraction approach the terminator plays
the role of the self-energy. As a result we clearly see the widening of the
defect structures with increasing disorder. No periodic supercell technique can give this disorder broadening
 with accuracy. 
 One added advantage of the real space approach  is that we can focus of the atom or cluster projected local density of states and assign different signatures in the spectrum to particular isolated defects or defect clusters.

We  propose that the real space recursion method is a powerful technique for the study of
extended defects in disordered solids. Our application to random Stone-Wells defects in graphene justifies our proposal.

\section*{Acknowledgements}
SC would like to thank DST, India for financial support through the Inspire Fellowship. This work was done under the HYDRA collaboration between our institutes.

\end{document}